\documentclass[a4paper,superscriptaddress, preprint,aip]{revtex4-1}

\usepackage{graphicx}
\usepackage{natbib}
\usepackage{amsmath}

\begin{document}
\title{Domain wall dynamics in a single CrO$_2$ grain}


\author{P. Das}
\email[]{das@physik.uni-frankfurt.de}
\affiliation{Institute of Physics, Goethe University, Max von Laue Str. 1, 60438 Frankfurt (M), Germany.}
\affiliation{Max Planck Institute for Chemical Physics of Solids, Noethnitzer Str. 40, 01187 Dresden, Germany}

\author{F. Porrati}
\affiliation{Institute of Physics, Goethe University, Max von Laue Str. 1, 60438 Frankfurt (M), Germany.}

\author{S. Wirth}
\affiliation{Max Planck Institute for Chemical Physics of Solids, Noethnitzer Str. 40, 01187 Dresden, Germany.}

\author{A. Bajpai}
\affiliation{Institute of Solid State Research, IFW-Dresden, 01171 Dresden, Germany.}

\author{M. Huth}
\affiliation{Institute of Physics, Goethe University, Max von Laue Str. 1, 60438 Frankfurt (M), Germany.}

\author{Y. Ohno}
\affiliation{Laboratory for Nanoelectronics and Spintronics, Research Institute of Electrical Communication, Tohoku University, Sendai 980-8577, Japan.}

\author{H. Ohno}
\affiliation{Laboratory for Nanoelectronics and Spintronics, Research Institute of Electrical Communication, Tohoku University, Sendai 980-8577, Japan.}

\author{J. M\"uller}
\affiliation{Institute of Physics, Goethe University, Max von Laue Str. 1, 60438 Frankfurt (M), Germany.}



\begin{abstract}
Recently we have reported on the magnetization dynamics of a single CrO$_2$ grain studied by micro Hall magnetometry (P.\ Das et al., {\em Appl.\ Phys.\ Lett.} \textbf{97} 042507, 2010). For the external magnetic field applied along the grain's easy magnetization direction, the magnetization reversal takes place through a series of Barkhausen jumps. Supported by micromagnetic simulations, the ground state of the grain was found to correspond to a flux closure configuration with a single cross-tie domain wall. Here, we report an analysis of the Barkhausen jumps, which were observed in the hysteresis loops for the external field applied along both the easy and hard magnetization directions. We find that the magnetization reversal takes place through only a view configuration paths in the free-energy landscape, pointing to a high purity of the sample. The distinctly different statistics of the Barkhausen jumps for the two field directions is discussed.
\end{abstract}
\maketitle
\section{Introduction}

In the realm of magnetism, the concept of domains and domain walls (DWs) are at the core of the understanding of the spontaneous magnetization of a ferromagnetic material. The domains, which are formed as a result of minimization of the total free energy, undergo structural changes under the influence of an external field. In real materials, the structural or compositional imperfections lead to several minima in the free energy landscape. These imperfections act as pinning centers for the DWs. An understanding of the dynamics of DWs in the presence of pinning centers is critical for the understanding of parameters like coercivity and remanence, which is essential for improving materials for practical applications. However, this is often difficult to realize in real samples consisting of several domains and DWs, where one measures an average magnetic signal and the details of the configuration path followed by the various DWs (due to local effects of pinning centers) are averaged out. Therefore, in order to gain detailed and quantitative information on the magnetization processes, as e.g.\ the pinning center distribution in a sample, it is essential to investigate samples where a single DW is present.

In literature, several studies on the dynamics of a single DW have been reported where, in most cases, a single wall is artificially created \cite{AtkinsonNatureMat2003, BedauPRL2008, HayashiPRL2006}. In the present study, we explore the possibilities of quantitative investigation of the dynamics of a single DW in an individual CrO$_2$ micro-grain of dimensions of approximately $5\,\times\,1.2\,\times\,0.7\,\mu$m$^3$. The DW in the present case is not artificially created (see below). CrO$_2$ has long been known as a material used for information storage in magnetic tapes. However, this half-metallic ferromagnet with a $T_C$ above room temperature ($\sim 395$\,K) attracted a renewed interest because of its very high spin polarization ($\sim\,100\%$), a property which is interesting for spintronics applications \cite{CoeyJAP2002}. Our magnetic studies of a single CrO$_2$ grain indicate that the ground state of the grain corresponds to a closure domain pattern with a single cross-tie DW \cite{DasAPL0972010}. We observe that upon application of an external field $\vec{H}_{\rm ext}$ along the long axis of the grain, which is the easy magnetization axis (EMD), the magnetization reverses due to the motion of the cross-tie DW. In this paper, we show that when the external field is applied along the hard magnetization direction (HMD) of the grain, the reversal involves the motion of the 90$^\circ$ DWs at the end caps. Accordingly, the histograms of the magnitudes of Barkhausen jumps strongly differ for the different field directions.

\section{Experimental}

High purity CrO$_2$ samples for the present studies were synthesized in a two step process. As a first step, intermediate precursor oxide is prepared by heating CrO$_3$ flakes or granules and then pellets formed from the precursor undergo a further heat treatment at appropriate temperature and at ambient pressure. The method is successfully used to tune the parameters like grain size and the grain boundary density. The grains thus prepared are of rod like shape. The details of the synthesis is described in Ref.\,\cite{BajpaiAPL2005, BajpaiPatent}.

\begin{figure}[h]
\includegraphics[width=20pc]{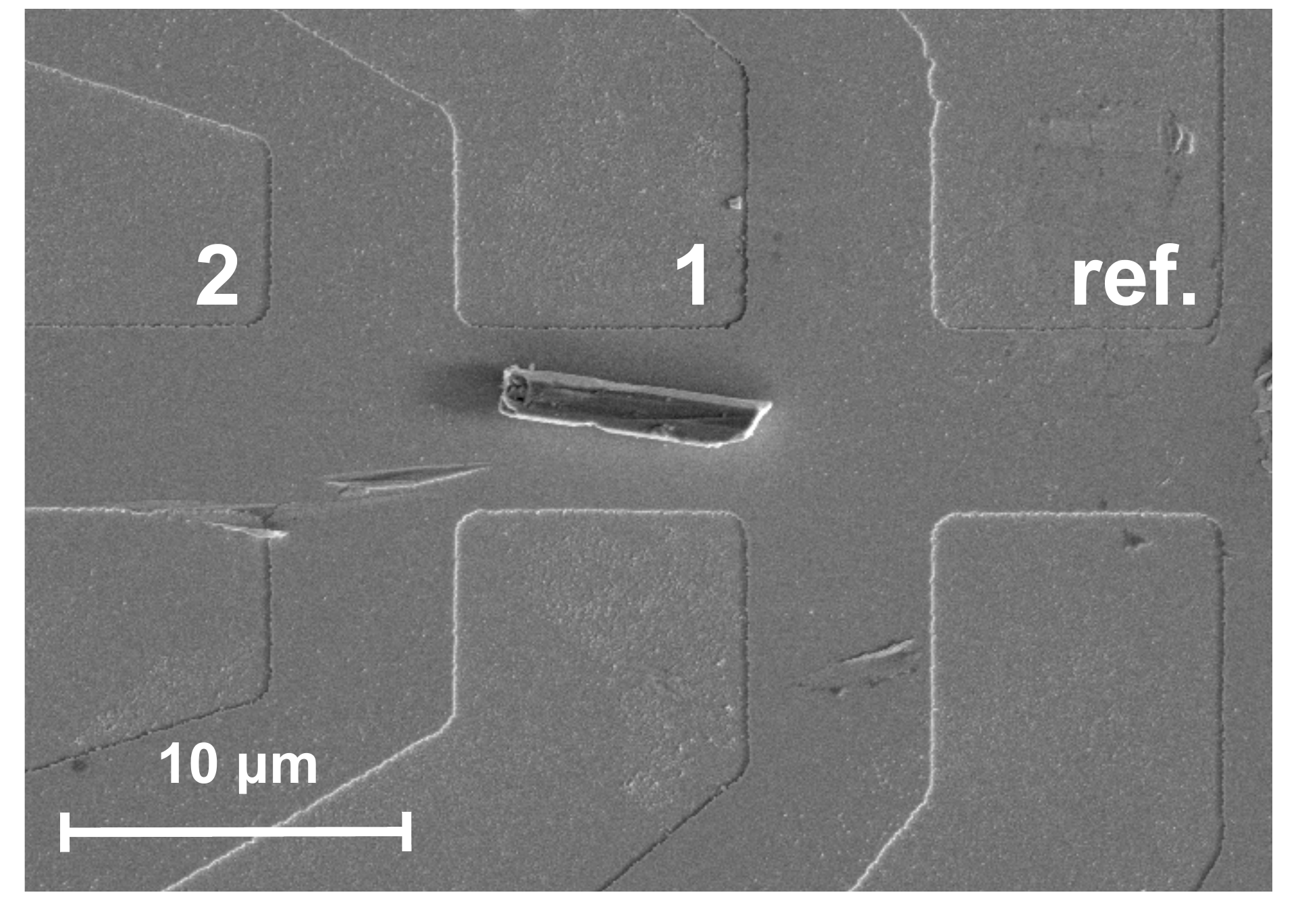}\hspace{5pc}%
\begin{minipage}[b]{10pc}
\caption{\label{Fig_Sample}SEM image of a CrO$_2$ grain placed on a Hall device. Note that the position of the grain is slightly misaligned ($\sim 7^\circ$) with respect to the two Hall crosses. A reference cross is used for the background correction (see text).}
\end{minipage}
\end{figure}

In order to study the magnetization dynamics in a single grain, we employed a micro-Hall magnetometer based on two-dimensional electron gas (2DEG) that forms at the interface of high-mobility material GaAs/AlGaAs heterostructures \cite{KentJAP1994}. Hall crosses with active areas of $5 \times 5\,\mu$m$^2$ were prepared using standard photolithography followed by wet etching technique. A single grain of CrO$_2$ obtained by crushing a pellet was then picked and placed between two Hall crosses (1 and 2, see Fig\,\ref{Fig_Sample}) aiming to simultaneously measure the Hall signals due to the stray fields emanating from the two grain ends. The 2DEG is sensitive to the perpendicular component of the average stray field $\langle B_z \rangle$. The measured Hall voltage is converted to $\langle B_z \rangle$ using the Hall coefficient of the device measured at the corresponding temperature.


\section{Results and discussions}

\begin{figure}[h]
\includegraphics[width=38pc]{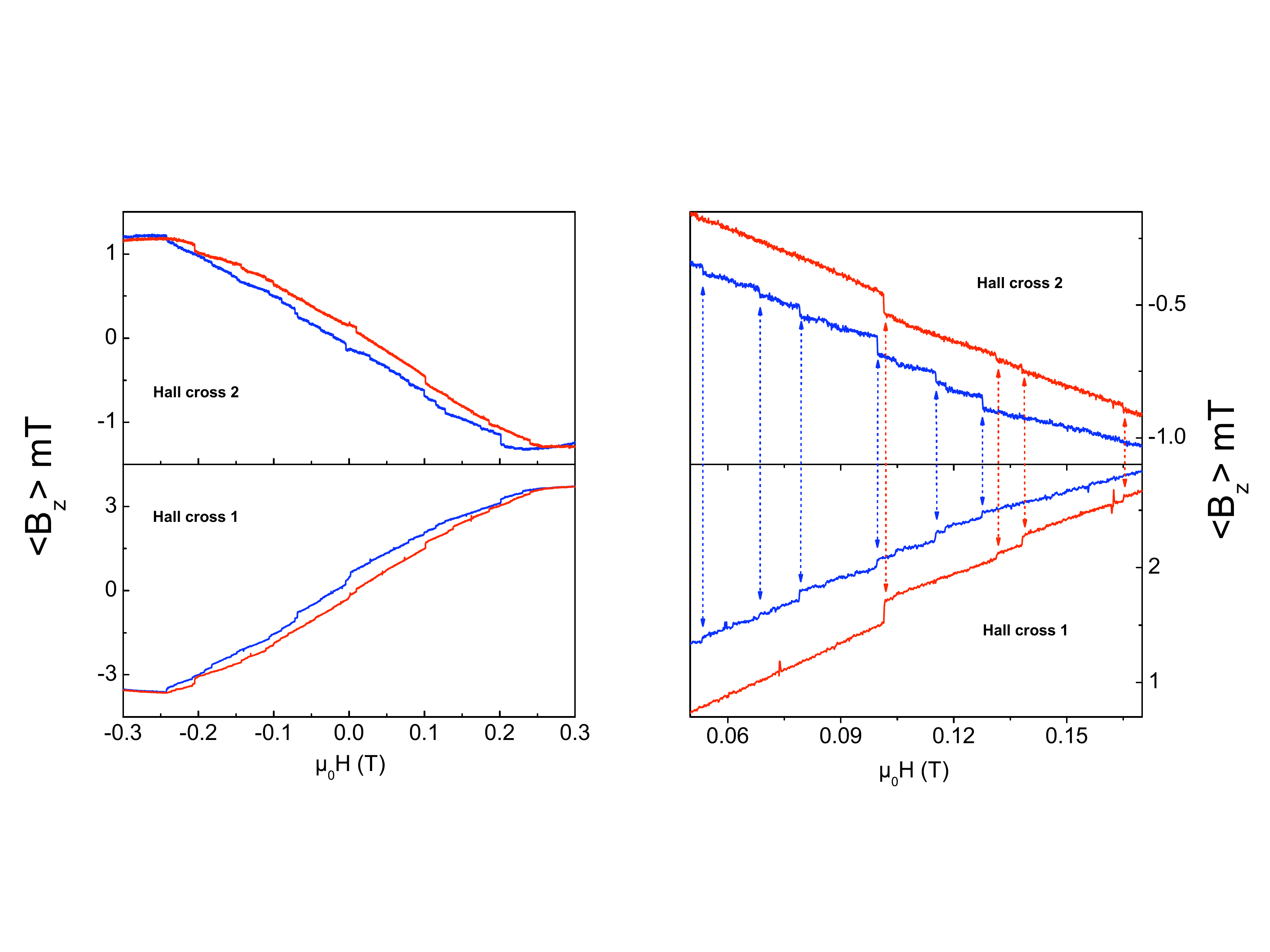}%
\caption{\label{Fig_hysteresis}(left) Hysteresis loops measured at $T$\,=\,5\,K with $\vec{H}_{\rm ext}$ applied along one of the HMD (field in the plane of the Hall device, cf.\ Fig.\,\ref{Fig_Sample}). Zoomed-in figure (right) show the sharp correlated Barkhausen jumps at two grain ends. }

\end{figure}

\begin{figure}[h]
\includegraphics[width=30pc]{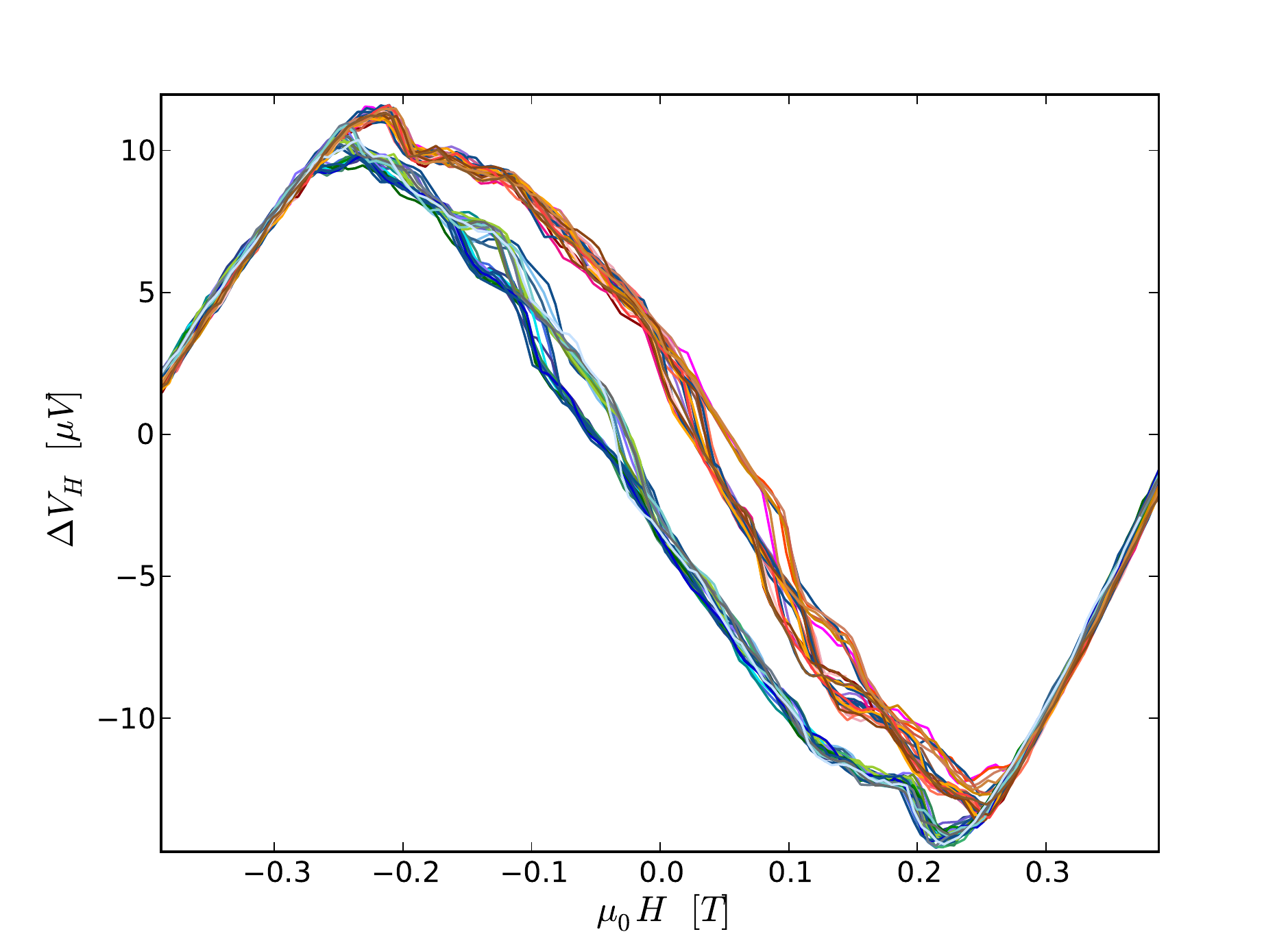}\hspace{2pc}%
\begin{minipage}[b]{38pc}\caption{\label{Fig_hysteresis50} 30 out of 50 successively measured hysteresis loops (gradiommetry setup, see text) with the applied field along one of the hard directions. The data are shown without background correction.}
\end{minipage}
\end{figure}

In order to obtain quantitative information on the magnetization dynamics of the grain, the Hall voltages at both grain ends were simultaneously measured while sweeping the external magnetic field $\vec{H}_{\rm ext}$ applied perpendicular and parallel to the long axis of the grain. As mentioned above, the EMD of the grain lies along the long axis, which is the [001] direction. The changes in the stray field were determined from the measured Hall signals after removing the background (due to a small misalignment of the device with respect to $\vec{H}_{\rm ext}$), which was measured at an empty reference cross. Figure\,\ref{Fig_hysteresis} shows the hysteresis loops measured in $\langle B_z \rangle$ versus $H_{\rm ext}$ with $\vec{H}_{\rm ext}$
 applied along a HMD. The two loops correspond to the signals measured simultaneously at the two ends of the grain, denoted by 1 and 2, see Fig.\,\ref{Fig_Sample}. For these measurements, a slow continuous field sweep was applied such that a complete hysteresis loop was measured in $\sim 22$\,hrs. Such a slow sweep rate allows to track and measure the DW motion in details. The magnetization saturates at the anisotropy field $\mu_0 H_{\textrm{a}} \sim 250$\,mT. Using this, we obtain an effective uniaxial anisotropy constant of $K_{\rm eff}\sim\, 7.3\times 10^4$ J/m$^3$ 
from which a uniaxial magnetocrystalline anisotropy constant $K_1$ $\sim$\,1.3\,$\times$\,10$^4$\,J/m$^3$ follows (see ref.\,\cite{DasAPL0972010} for details).
From Fig.\,\ref{Fig_hysteresis}, we make the following observations: First, the magnetic signal at grain end 1 (i.e.\ measured at Hall cross 1) is larger and opposite in sign to that measured at Hall cross 2. The larger signal at Hall cross 1 is due to the asymmetry in the placement of the grain with respect to the crosses (see Fig.\,\ref{Fig_Sample}).
Secondly, the magnetization reversal is accompanied by a series of sharp Barkhausen jumps. Interestingly, we find that the jumps at both grain ends occur simultaneously as is evident from Fig.\,\ref{Fig_hysteresis}. Correlated jumps were also observed, when the external field applied was applied along the EMD \cite{DasAPL0972010}. These simultaneous jumps in magnetic signal from both ends of the grain are consistent with a single DW that is being pinned/depinned by the force exerted by the field sweep. In order to find out about the variety of paths in configuration space of the free-energy landscape, a series of 50 successive hysteresis loops, shown in Fig.\,\ref{Fig_hysteresis50}, has been measured. The set of data demonstrates that the magnetization reversal takes place along only a few ($2 - 3$) distinct, stochastically chosen paths corresponding to different configurations of the energy landscape. These measurements have been carried out in a gradiometry setup where the Hall voltage at the cross carrying the sample is compensated by applying a current of same magnitude but opposite sign through an empty reference cross. This results in a differential signal $\Delta V_H$ solely due to the sample's magnetization \cite{WirthPRB632001,LiPRB2005}. The data in Fig.\,\ref{Fig_hysteresis50} are shown without background correction.
The data suggest the motion of a single DW, since for multiple walls the magnetization reversal wouldn't be expected to follow only a few distinct paths. This finding was confirmed by micromagnetic simulations, which revealed two different field regimes of the grain's magnetization. The ground state corresponds to a closure domain configuration (labeled as configuration A) with a single cross-tie DW parallel to the long axis of the grain. However, there exists a metastable state (configuration B) with three domains, which is slightly higher in energy. Application of a small field along the direction opposite to that of the main domain brings the system back to the lowest energy state. We refer to Ref.\,\cite{DasAPL0972010} for the details on the domain configurations. As mentioned above, we observed Barkhausen jumps simultaneously at both grain ends for fields applied both along the EMD and HMD. For  $\vec{H}_{\rm ext} \parallel$ EMD, the magnetization reversal takes place through DW motion in a direction perpendicular to the direction of the applied field. For perpendicular fields smaller than the anisotropy field $H_{\textrm{a}}$ most likely the metastable three-domain configuration B is formed. At $\mu_0 \vec{H}_{\rm ext} \approx \pm 200$\,mT we observe downward jumps (see Figs.\,\ref{Fig_hysteresis} and \ref{Fig_hysteresis50}) in the magnetic signal where the single cross-tie DW with closure domain configuration A is formed. We relate this transition to the observed transition from configuration A to B (see Fig.\,3 in Ref.\,\cite{DasAPL0972010}) at $\mu_0 \vec{H}_{\rm ext} \approx \pm 20$\,mT for $\vec{H}_{\rm ext}$ along the EMD.

\begin{figure}[h]
\includegraphics[width=40pc]{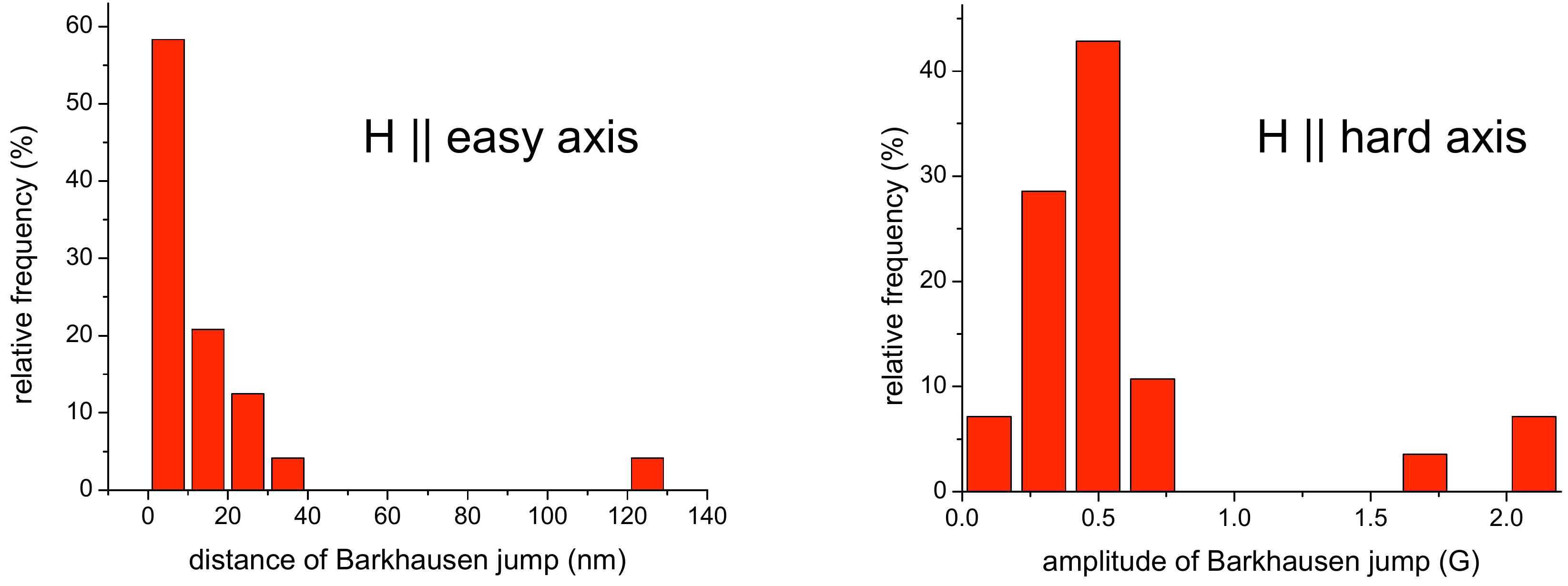}\hspace{2pc}%
\caption{\label{Fig_Hystogram}Histograms of Barkhausen jump heights observed for $\mu_0 \vec{H}_{\rm ext}$ along both EMD and HMD at $T = 5$\,K. The amplitude of Barkhausen jumps is directly proportional to the spatial distance traversed by the domain wall (for $\vec{H}_{\rm ext} \parallel$ EMD, see text and Ref.\cite{DasAPL0972010}). The relative frequency in the ordinate refers to the no. of jumps of certain size with respect to the total number of jumps. The smallest jump which was observed at $T$\,=\,5\,K is $\sim$\,0.8 nm.}
\end{figure}

In the present case where $\vec{H}_{\rm ext} \parallel$ HMD, the observation that the stray field changes simultaneously at both grain ends indicates that the motion of the 90$^\circ$ wall at one end influences that at the other end of the grain via the connecting cross-tie DW. Such a behavior would be expected in case of a rigid connecting wall. In order to understand this in more detail, we analyze the Barkhausen jumps observed for $\vec{H}_{\rm ext}$ along both the EMD and HMD. In Fig.\,\ref{Fig_Hystogram}, we show the histograms of the magnitude of the magnetization changes (i.e.\ size of the Barkhausen jumps) observed for both field directions at $T = 5$\,K. The behavior, which reflects the distribution of pinning centers is qualitatively different for the two field directions.
For $\vec{H}_{\rm ext} \parallel$ EMD the step heights can be directly converted to the distance ($\Delta x$) of DW motion between two pinning centers using $\Delta \phi = 2 M_s d
\Delta x$, where $\phi \propto \langle B_z \rangle$ is the flux
emanating from the grain of thickness $d$, and $M_s$ is the saturation magnetization, see Ref.\,\cite{DasAPL0972010} for details.
For $\vec{H}_{\rm ext} \parallel$ EMD the data show an asymmetric distribution,
whereas for $\vec{H}_{\rm ext} \perp$ EMD the distribution of the magnitude of Barkhausen jumps is more symmetric with mean values of the step heights of $\sim 107\,\mu$T in the former case and $\sim$ 65\,$\mu$T in the latter case, respectively. Jumps with much larger magnitudes correspond to the switching between domain configurations A and B.
From the easy-axis measurements we roughly estimated a density of pinning centers of $1 - 10$\,$\mu$m$^{-3}$ implying a high purity of the investigated grain, in agreement with the small number of configuration paths as seen from Fig.\,\ref{Fig_hysteresis50}.
We find that about $1/3$ of the magnetization changes through DW pinning/depinning and about $2/3$ through free motion of DWs and magnetization rotation.
Different heights observed for the different field directions at the same temperature are in agreement with the assumption that the direction of DW motion is different for $\vec{H}_{\rm ext}$ along the EMD and HMD. For the latter case, the observed Barkhausen jumps involve the motion of the 90$^\circ$ DW being affected by pinning centers.
\section*{Conclusions}
We have studied the DW dynamics in a single CrO$_2$ grain using micro-Hall magnetometry. The motion of the DW was tracked and quantitative information on the pinning-center distribution was obtained. We find that the magnetization reversal takes place through only a view configuration paths in the free-energy landscape, pointing to a high purity of the sample. The observed Barkhausen jumps for the two field directions exhibit at distinctly different statistical behavior, since for the field along the hard axis the magnetization reversal involves the motion of the 90$^\circ$ DWs (forming closure domains). The data suggest that the cross-tie DW along the long axis of the grain is a rigid wall.





\begin{thebibliography}{99}



\bibitem{AtkinsonNatureMat2003}Atkinson D, Allwood D A, Xiong G, Cooke M D, Colm C F and Cowburn R P 2003 \textit{Naturematerials} \textbf{2} 85.
\bibitem{BedauPRL2008}Bedau D, Kl\"aui M, Hua M T, Kryzk S, R\"udiger U, Faini Gand Vila L 2008 \textit{Phys. Rev. Lett.} \textbf{101} 256602 (2008).
\bibitem{HayashiPRL2006}Hayashi M, Thomas L, Rettner C, Moriya R, Jjiang X and Parkin S S P 2006 \textit{Phys. Rev. Lett.} \textbf{97} 207205.
\bibitem{CoeyJAP2002}Coey J M D and Venkatesan M 2002 \textit{J. Appl. Phys.} \textbf{91} 8345.
\bibitem{DasAPL0972010} Das P, Porrati F, Wirth S, Bajpai A, Ohono Y, Ohono H and M\"uller J 2010 \textit{Appl. Phys. Lett.} \textbf{97} 042507.
\bibitem{BajpaiAPL2005} Bajpai A and Nigam A K 2005 \textit{Appl. Phys. Lett. } \textbf{87} 222502.
\bibitem{BajpaiPatent}Bajpai A and Nigam A K , US Patent 7276226.
\bibitem{KentJAP1994}Kent A D, von Moln\'ar S, Gider S and Awschalom D D 1994 \textit{ Journ. Appl. Phys.} \textbf{76} 6656.
\bibitem{WirthPRB632001} Wirth S, Anane A and von Moln\'ar S 2001 \textit{Phys. Rev. B} \textbf{63} 012402.
\bibitem{LiPRB2005}Li Y, Xiong P, von Moln\'ar S, Ohno Y and Ohno H 2005 \textit{Phys. Rev. B} \textbf{71} 214425.
\end{thebibliography}
\end{document}